\documentclass[reprint,amsmath,amssymb,aps,prd,showpacs,floatfix]{revtex4-1}
\usepackage{graphicx}

\newtheorem{definition}{Definition}

\begin{document}

\title{Causal discrete field theory for quantum gravity}

\author{K. V. Bayandin}

\date{\today}

\begin{abstract}

The proposed theory of causally structured discrete fields studies integer values on directed edges of a self-similar graph with a propagation rule, which we define as a set of valid combinations of integer values and edge directions around any vertex of the graph. There is an infinite countable number of variants of the theory for a given self-similar graph depending on the choice of propagation rules, some of these models can generate infinite uncountable sets of patterns. This theory takes minimum assumptions of causality, discreteness, locality, and determinism, as well as fundamental symmetries of isotropy, CPT invariance, and charge conservation. It combines the elements of cellular automata, causal sets, loop quantum gravity, and causal dynamical triangulations to become an excellent candidate to describe quantum gravity at the Planck scale. In addition to the self-consistent generation of spacetime and metrics to describe gravity and an expanding closed Universe, the theory allows for the many-worlds interpretation of quantum mechanics. We also demonstrate how to get to unitary evolution in Hilbert space for a stationary Universe with deterministic propagation.

\end{abstract}

\pacs{04.60.-m, 04.60.Nc, 04.90.+e, 03.65.-w, 03.65.Ta, 03.65.Ud}

\maketitle

\section{Introduction}

There is a long-standing idea that the Universe must be deterministic, discrete, and local at the Planck scale~\cite{tHooft2005}. There is also a temptation to build such a Universe via cellular automata that can demonstrate complex behavior even from simple rules~\cite{Conway1970}, but the patterns from cellular automata have a global time axis, which is not applicable to describe general relativity. Another idea is that the Universe may be a tiling of spacetime~\cite{Penrose1989}. We introduce a theory that combines these two ideas into causal tilings of spacetime that represent patterns of integer values on edges of self-similar directed acyclic graphs.

On a historical note, already at the time of the foundation of quantum mechanics and general relativity, there was a suggestion that the Universe is a computation~\cite{Zuse1967}. Later, several authors discussed this idea further~\cite{Fredkin2005},~\cite{Schmidhuber1997},~\cite{Wolfram2002}. The causal set theory was the first to use the idea of causality to study the structure of spacetime at the Planck scale~\cite{Sorkin2005}, but it required external dynamics to reproduce physically meaningful results. Casual dynamical triangulation allows to create physically meaningful causal tilings of spacetime~\cite{Loll2004}, but it required an assumption for the Hilbert-Einstein action with model parameters $\alpha$ and $\Lambda$. One of the closest predecessors to the proposed theory is loop quantum gravity and its representation with spin newtworks~\cite{Smolin1988},~\cite{Smolin1990}. However, this theory again required external assumptions and started from a formulation of general relativity with new parameters. The approach of this paper is to start purely from discrete geometry and to search for ways how general relativity and quantum mechanics emerge from there.

We organize the paper in the following way. First, we introduce a path-integral approach for a general case of discrete integer fields on edges of a self-similar graph (lattice) in any dimension. We postulate discrete fields as integer values and edge directions of the graph. We introduce a propagation rule that describes all valid combinations of fields on edges around any vertex of the graph. We allow only for a single initial vertex.

Second, we introduce a toy model in two-dimensions on a honeycomb graph to show the intuition behind the theory. We use a Matlab code that generates and visualizes two-dimensional patterns on this graph~\cite{matlab}. We show examples of patterns that are generated by this code throughout the paper in order to support the discussion.

Third, we discuss the phenomenology that can be demonstrated by this theory. We describe a family of generalized diamond crystal lattices that allow for isotropic pattern generation, which is crucial for the representation of a uniform space. We describe the Big Bang that originates causally from an initial vertex. We demonstrate the existence of black holes at the Planck scale. We introduce the future and past light cones. We propose a covariant definition of a quantized spacetime interval purely from geometry. This measure adds scale to the causal relationship between vertices of the graph and therefore allows to reconstruct a continuous Riemannian manifold. We postulate the energy-momentum tensor together with the cosmological term $\Lambda$ from the Einstein field equations. We introduce CPT invariance and charge conservation within the theory. We make a hypothesis that elementary particles and low energy quantum field theories may arise from the dynamics of the patterns on the graph. We suggest that non-computable charge conserving propagation rules may be relevant for the creation of mass and existence of inertia. We demonstrate the way to calculate entropy for a case of a stationary Universe. We briefly discuss the relevance of this theory to describe quantum phenomena such as Heisenberg uncertainty, the double-slit experiment, EPR paradox, Bell's theorem, and quantum entanglement. For this matter, we refer to the many-worlds interpretation of quantum mechanics. Finally, we show a way to get to unitary evolution in a Hilbert space in the case of a stationary Universe with deterministic propagation.

We conclude with a discussion of possible ways to test the theory in order to make it falsifiable.

\section{General theory with path-integral formulation}

The proposed theory describes all valid configurations of integer values and edge directions on self-similar graphs (lattices) given a propagation rule. We start with important definitions:

\begin{definition}
A translation of a graph is a permutation of its vertices that preserves connectedness between all vertices of the graph.
\end{definition}

\begin{definition}
A self-similar graph has all vertices with $k$ direct neighbors, and it allows for a translation that places any vertex $i$ instead of any other vertex $j$.
\end{definition}

\begin{definition}
Any edge of a self-similar graph has a class (angular direction) assigned to it: $c=1,\dots,k$, so that any translation does not mix classes of edges and moves all edges of one class to another class.
\end{definition}

An example of a self-similar graph is a cubic lattice in n-dimensions. There are $k=2d$ direct neighbors for each vertex and $c=2d$ classes of edges two for each direction along each dimension $d=n$. Cubic lattices are simple and easy for modeling, but they may be irrelevant for not having enough isotropy.

Another example is a general diamond crystal lattice in n-dimensions, which is a subset of vertices of an (n+1)-dimensional cubic lattice with integer coordinates that sum either to one or zero. The edges for this graph are formed between direct neighbors in the initial (n+1)-dimensional cubic lattice. A general diamond crystal lattice in $d=n$ dimensions has $k=d+1$ direct neighbors for each vertex and $c=d+1$ classes of edges one for each direction of edges. A general diamond crystal lattice is a honeycomb graph in two dimensions and a diamond crystal lattice in three dimensions.

\begin{definition}
An integer field variable $\psi_{ij}=-m,\dots,-1,0,1,\dots,m$ is defined for any two adjacent vertices $i$ and $j$, where a positive integer value means direction from $i$ to $j$, and a negative integer value means direction from $j$ to $i$. These fields are antisymmetric: $\psi_{ij}=-\psi_{ji}$.
\end{definition}

Zero value of $\psi_{ij}$ means that the edge does not have an assigned integer value and direction. Some variations of the theory may forbid zero integers on edges. In this case, all edges will have an assigned direction.

\begin{definition}
Any vertex $i$ has a set of interger fields on its edges that are defined by a tuple $\mathcal{T}_{i}=\{\psi_{i,v(i,1)},\dots,\psi_{i,v(i,k)}\}$, with a vicinity function $j=v(i,c)$ that orders edges $\psi_{ij}$ by classes $c=1,\dots,k$.
\end{definition}

\begin{definition}
A propagation rule $\mathcal{P}$ is defined for any vertex $i$ as a subset of all valid tuples of integer fields on edges: $\mathcal{P}\subset\forall\{\psi_{i,v(i,1)},\dots,\psi_{i,v(i,k)}\}$.
\end{definition}

\begin{definition}
A vertex $i$ satisfies a propogation rule $\mathcal{P}$ if the tuple of integer fields around the vertex $\mathcal{T}_{i}$ belongs to the subset of all allowed tuples of integer fields for the propagation rule: $\mathcal{T}_{i}\in\mathcal{P}$.
\end{definition}

\begin{definition}
We introduce three causality constraints for the discrete fields $\psi_{ij}$:
\begin{itemize}
    \item Propagation rule: $\forall{i}:\mathcal{T}_{i}\in\mathcal{P}$
    \item No cycles: $\nexists{i_1,\dots,i_n}:\psi_{i_1i_2}>0,\dots,\psi_{i_ni_1}>0$
    \item Unique initial vertex: $\exists! b : \forall{j}\ \psi_{bj}>0$
\end{itemize}
\end{definition}

These three constraints make sure that the edges of the graph form a directed acyclic graph (DAG) and all vertices of the graph satisfy a specific propagation rule $\mathcal{P}$.

Now we can formulate the main idea of the proposed theory:

\begin{quote}
The casual discrete field theory (CDFT) studies properties of integer value fields $\psi_{ij}$ on edges of self-similar graphs under causality constraints with a given propagation rule $\mathcal{P}$.
\end{quote}

We define any measurable physical quantity as a functional over field configurations: $F[\psi]$. Also for convenience, we define a notation for a constraint functional $C_{\mathcal{P},b}[\psi]$ for a certain propagation rule $\mathcal{P}$ and an initial vertex b. This constraint functional is proportional to the likelihood of each configuration of fields that satisfy causality constraints and $0$ otherwise. This constraint functional $C_{\mathcal{P},b}[\psi]$ is a discrete analog of the action term $e^{iS[\psi]}$ and time ordering for path integrals in quantum field theories.

We define an expectation value of an operator $F$ with an analog of path integral summation over all valid field configurations undger causality constraints:
\begin{equation} \label{eq:path_integral}
\langle F\rangle=\frac{\int\mathcal{D}\psi F[\psi] C_{\mathcal{P},b}[\psi]}{\int\mathcal{D}\psi C_{\mathcal{P},b}[\psi]}
\end{equation}

The theory is constructed with self-similar graphs because they are composed of vertices with identical local topology. Any vertex can be replaced by any other vertex via translations so that the graph would entirely repeat itself. The theory can be easily generalized even further by allowing lattices with several classes of vertices with different local topology and different propagation rules for each class, but we keep it out of the discussion for simplicity.

A propagation rule for a given self-similar graph is the cornerstone of the proposed theory because it fully defines all patterns of integer fields and edge directions. A propagation rule is a discrete analog of interactions between fields in continuous field theories.

\begin{definition}
Any propagation rule $\mathcal{P}$ for a self-similar graph with $k$ nearest neighbors for each vertex is encoded by a pair of numbers $m:E$, where $m$ is a maximum integer value on edges and $E$ is an encoding of a subset of all numbers in (2m+1)-base numeric system with $k$ digits. Thus the encoding $E$ is a number from $0$ to $2^{(2m+1)^k}-1$.
\end{definition}

Encoding of all propagation rules with two positive integer numbers means that there is an infinite countable number of different models for a given self-similar graph.

We use integer values on edges to keep the theory discrete with countable variables, but the most general at the same time. We use a limit of $m$ for these values to avoid infinite parameter space for propagation rules. A more general theory is possible with infinite countable or continuous sets of values on edges with some functional structure for propagation rules, but this goes beyond the scope of this paper.

We have defined the most general propagation rule $\mathcal{P}$. This propagation rule may have additional limitations and symmetries like the absence of zero value edges, isotropy, CPT invariance, and charge conservation. We consider these constraints in more detail in the discussion section.

A specific propagation rule $\mathcal{P}$ for a given self-similar graph defines a CDFT model. Practical calculations show that even models with small $m$ can generate complex causally related patterns or, in other words, tilings of space.

\begin{definition}
Each valid configuration of integer fields on edges corresponds to a tiling of space, where each tile corresponds to a vertex with a particular tuple $\mathcal{T}_{i}$ from valid tuples in the propagation rule $\mathcal{P}$. In this tiling two tiles $\mathcal{T}_{i}$ and $\mathcal{T}_{j}$ may be adjacent only if they share an edge with the same integer value $\psi_{ij} = -\psi_{ji}$. 
\end{definition}

From the perspective of tilings, a propagation rule is a set of valid tiles that can tile the whole space, and any two adjacent tiles have to have matching absolute values of positive and negative integers from $1$ to $m$.

The presence of the initial vertex $b$ is essential to define a starting point for patterns on the graph, whereas multiple initial vertices will add too many degrees of freedom in this theory. There is also a sequential growth interpretation for patterns on the graph. Each field configuration on a graph allows at least one sequential enumeration of vertices so that each new vertex has all its directed edges from already enumerated adjacent vertices.

Some propagation rules allow for the computation of patterns by sequential growth of tiles one by one with all valid local combinations. These propagation rules may generate only periodic patterns or may allow for more complex patterns. Also, there are remaining propagation rules that forbid certain combinations of values on incoming edges. In this situation, a sequential placement of tiles may end up with a stalemate when there is no valid tile for the next vertex. The propagation rules of the second type may not have a possible tiling at all, or they may generate non-computable tilings in a sense defined by Roger Penrose~\cite{Penrose1989}, who had ideas related to the proposed theory, but his famous tilings lacked causal structure which is essential for the CDFT.

\begin{definition}
A propagation rule is trivial if it generates only periodic tilings or does not generate any tiling at all.
\end{definition}

\begin{definition}
A propagation rule is computable if it allows for a deterministic construction of tilings by the sequential placing of tiles one by one in a recursion by vertices starting from the initial vertex $b$ and by choosing from all valid tiles for each next vertex.
\end{definition}

Non-trivial computational propagation rules are useful for studying because they allow for the computational generation of patterns and calculation of desired metrics and properties. However, it is highly likely that non-trivial non-computational propagation rules are more relevant for a description of physical reality, because they intrinsically non-local globally: a local change in a type of tile for a vertex $i$ may put constraints on a type of tile for another vertex $j$ that does not have a direct causal relationship with the vertex $i$. This situation resembles quantum non-locality, which we cover in more detail in the discussion section.

All possible patterns for a given graph and a given propagation rule form a set of overlapping patterns. Each pair of patterns in this set have ``joint past", which is a maximum number of connected vertices identical for both patterns. As a result, this set defines all possible evolutions for a given variant of a causal discrete field theory. We call this set of patterns a Multiverse, which contains all copies of possible Universes. We postulate that each copy of a Universe participates in path integral summation over field configurations in Eq.~\ref{eq:path_integral}.

There are propagation rules that can generate an infinite uncountable number of Universes within a Multiverse covering a continuous spectrum of possibilities. We can demonstrate that for a simple propagation rule with $m=2$ and all possible incoming and outgoing directions of edges. Even for each fixed directed acyclic graph, there is an uncountable number of patterns of integer values of $1$ or $2$ on edges, because this corresponds to all possible subsets of an infinite countable set of directed edges.

The proposed CDFT program for future research is to study the properties of non-trivial patterns for different self-similar graphs and propagation rules. In the remaining part of this paper, we discuss why CDFT models are relevant for the description of physical reality.

\section{A two-dimensional example for illustration}

\begin{figure}
\begin{center}
\includegraphics[width=0.85\linewidth]{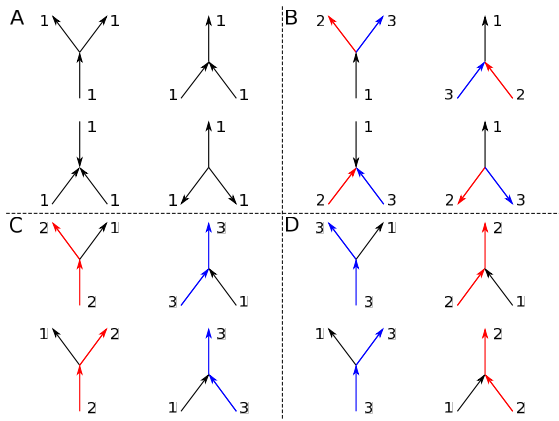}
\end{center}
\caption{An example of a propagation rule for two simple models on a honeycomb graph in two dimensions. Model A: four valid types of vertices in sector A. Model A-D: sixteen valid types of vertices in sectors A-D. Each type of vertex also allows all possible rotations. Edges with the same integer value have the same color: 1 - black, 2 - red, 3 - blue.}
\label{fig:FIG_1}
\end{figure}

We use two simple models in order to visualize the results of the discussion. We take a hexagonal honeycomb graph in two dimensions. The first model is the most simple one. It has only one integer $1$ on the edges of the graph and all four valid types of vertices with all possible directions of edges including all rotations, see Model A on Figure~\ref{fig:FIG_1}. The second model is a little bit more complicated. It has three integers $1-3$ on the edges of the graph and sixteen valid types of vertices including all rotations, see Model A-D on Figure~\ref{fig:FIG_1}.

An expample of enconding $m:E$ for the simplest possible Model A is $1:85197125$, where the second number equals to $\sum{2^i}$, where $i\in[0,2,2*3,2*3+2,2*9,2*9+2,2*9+2*3,2*9+2*3+2]$.

\begin{figure}
\begin{center}
\includegraphics[width=1.0\linewidth]{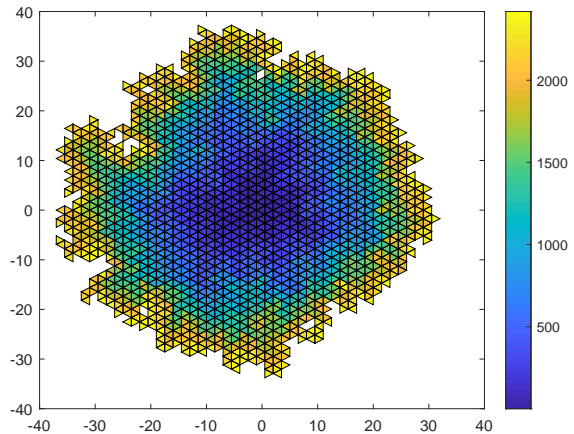}
\end{center}
\caption{A pattern on a honeycomb graph with Model A. The chart shows the result of propagation when the first tile have reached a vertex 50 edges away from the initial vertex $b$. Each triangle tile corresponds to the nearest neighbourhood of a vertex. Colors show the order of propagation.}
\label{fig:FIG_2}
\end{figure}

Model A is non-trivial and computable because it allows for all possible directions for a directed acyclic graph. We use a Matlab code to generate pattern evolution for Model A~\cite{matlab}. We start from the initial vertex $b$ and sequentially add vertices with valid types one by one. We pick each next vertex randomly from all possible locations and valid vertex types with equal likelihood. Figure~\ref{fig:FIG_2} shows an example for several first steps of evolution for Model A with vertices not far than 50 links from the initial vertex $b$.

Model A-D is also non-trivial because it contains all valid vertices of Model A. Model A-D is non-computable because it does not allow all possible combinations for incoming edges. The reason for having Model A-D as an additional example will be apparent during the discussion of symmetries for propagation rules.

\section{Discussion}

\textbf{Planck element of spacetime.} We now switch from pure mathematics to the discussion of physical phenomena that may be relevant for the proposed theory. The essential foundation of it is the concept of causality and discreteness of spacetime. We start with several essential definitions.

\begin{definition}
Each vertex $v$ of the graph is postulated to be an atomic Planck element of spacetime. Later on, these vertices are called spacetime events.
\end{definition}

\begin{definition}
A chain of links $l(v,w)$ between two events $v$ and $w$ is any connected sequence of edges of the graph regardless of their direction.
\end{definition}

\begin{definition}
A path of links $l_{\pm}(v,w)$ between two events $v$ and $w$ is a chain with all its edges directed from $v$ to $w$ for $l_{+}(v,w)$ or vice versa from $w$ to $v$ for $l_{-}(v,w)$.
\end{definition}

\begin{definition}
The length of a chain $|l(v,w)|$ or a path $|l_{\pm}(v,w)|$ is defined as a number of links in it.
\end{definition}

\begin{definition}
Every vertex $v$ has its past $w \in J^{-}(v): \exists l_{-}(v,w)$ and its future $w \in J^{+}(v): \exists l_{+}(v,w)$.
\end{definition}

\begin{figure}
\begin{center}
\includegraphics[width=1.0\linewidth]{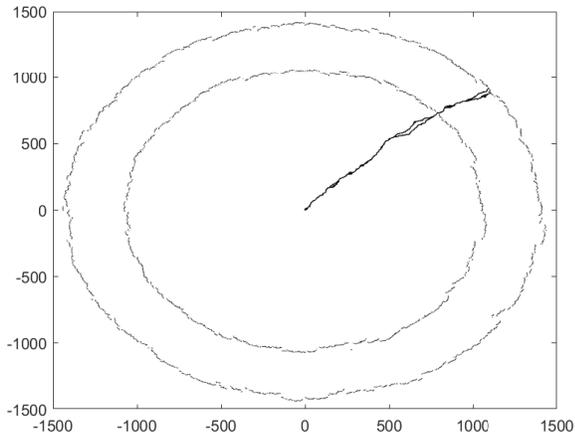}
\end{center}
\caption{An example of two epoches with $n=1500$ and $n=2000$ (circle tiles) with future and past light cones starting from an event from the first ephoch (triangle tiles). An evolution is shown for Model A only for events with causal descendants at epoch $n=2500$ (outside of interiors of black holes).}
\label{fig:FIG_3}
\end{figure}

These two sets of events are analogs of light cones originating from an event $v$ into the past and future, see Figure~\ref{fig:FIG_3}.

\textbf{The size of the Universe.} For each event $v$ we define a current size of the Universe, which does not have physical meaning, but essential for the following analysis.

\begin{definition}
A cosmological epoch is a set of events with equal minimal paths from the initial event $b$: $w\in E(n): \min(|l_{+}(b,w)|) = n$.
\end{definition}

By definition, any two events from the same epoch have at least the initial event $b$ as a joint predecessor. The number of events in a certain epoch defines the volume of the Universe at this stage. See Figure~\ref{fig:FIG_3} for an example of two epochs at $n=1500$ and $n=2000$ for Model A.

\textbf{The Big Bang} naturally occurs for non-trivial CDFT models as events of the future epochs expand from the initial event $b$. There is no global time in the model, which is a desired property for a covariant theory of general relativity.

The same propagation rule defines the vicinity of any event, so the Big Bang is happening at every event, which follows the ideas of Andrei Linde on the inflationary model of the Universe~\cite{Linde1982}. The proposed theory describes an eternally expanding closed Universe with a possibility that some regions collapse into giant gravitational singularities, so that local observers may perceive them as closed Universes with a finite lifetime.

\textbf{Black holes.} Some of the events may not have causal descendants at all. Suppose a vertex $i$ has only incoming links $\forall j:\psi_{ij}<0$. Some of the causal predecessors of this event $i$ also do not have descendants further, so all these events is an analog of an interior of a black hole at the Planck scale. The event horizon for such a black hole is a set of links that separate events without descendants further than the event $i$ from events with descendants further than the event $i$.

The event horizon is not a special object in this theory, and there is no a ``firewall". The paths of links that cross this border do not experience any change in local dynamics. The only consequence is that these paths will never reach infinity and they will collapse into a singularity at the center of a black hole after a finite number of steps.

\begin{figure}
\begin{center}
\includegraphics[width=1.0\linewidth]{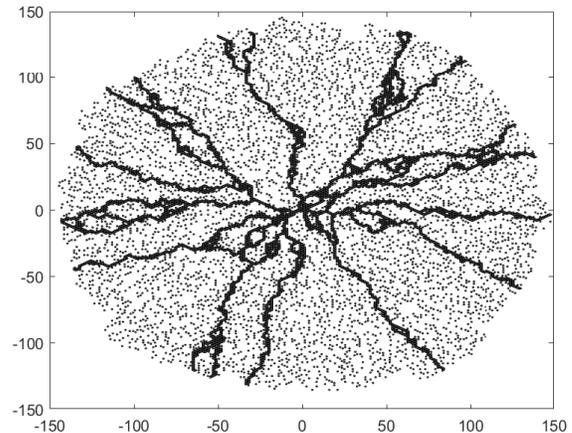}
\end{center}
\caption{A demonstration of black hole singularities. Evolution is shown for Model A. Propagation is shown for 200 epochs for events (triangle tiles) with causal descendents at epoch 2500 (outside of interiors of black holes). Circles show singularity events with no causal descendents.}
\label{fig:FIG_4}
\end{figure}

The proposed theory generates many singularity events, see Figure~\ref{fig:FIG_4} for example. Spacetime at the Planck scale forms quantum foam with high fluctuations of physical quantities and high density of the Planck scale black holes. In this example, propagation that reaches infinity concentrates within a few narrow bands of events that are separated by pockets of interiors of large-scale black holes. In four dimensions, the events that reach infinity are separated by four-dimensional pockets correspondingly. The cosmological inflation of the Universe should keep these narrow bands long enough for the size of the Universe to grow much faster than it expands.   

\textbf{Isotropy of space.} The real Universe demonstrates the isotropy of space so that all directions are equivalent. Any good theory on a lattice has to demonstrate this property too. There are for sure graphs and propagation rules in the proposed theory that do not have isotropic expansion from the initial event $b$. For example, in the case of a simple cubic lattice in n-dimensions, expansion with any propagation rule seems to preserve the shape of a cube with softened edges.

General diamond crystal lattices are more complex than, e.g., n-dimensional cubic lattices, but they are highly isotropic and have closer symmetry to a uniform space~\cite{Eppstein2009}. These lattices are unique because they have the minimum number of edges for each vertex, and these edges stretch uniformly in an n-dimensional space that embeds the lattice. Also, there is a translation of the graph that can exchange every two edges from the same vertex, which is not the case for a cubic lattice. Every vertex in a general diamond crystal lattice is a center of an n-dimensional simplex. The honeycomb graph is a two-dimensional case of a general diamond crystal lattice with a triangle simplex.

The propagation rule also has to be isotropic, so that local topology of edges around vertices does not depend on the direction of a class of edges.

\begin{definition}
A propagation rule $\mathcal{P}$ is isotropic, if for each tuple $\mathcal{T}_{i}=\{\psi_{i,v(i,1)},\dots,\psi_{i,v(i,k)}\}$ it contains all its rotations with respect to directions for each edge class. For a general diamond lattice the rotations of a tuple mean all parity preserving permutations of edge classes within the tuple. 
\end{definition}

There are at least two ways to check isotropy of patterns generated by a CDFT model. The first approach is to check numerically that the fluctuations of $\delta r /r$ are vanishing for epochs with large $n$. Here we take $r$ as a Euclidean distance between a vertex of an epoch and the initial vertex $b$ in an n-dimensional Euclidean space that embeds the lattice. We mean that a lattice is embedded into a Euclidean space of a minimum dimension when the distance between adjacent vertices is equal to one, and the angles between any two pairs of edges are the same if there is a translation of the graph that exchanges these pairs of edges. The second approach purely follows from the structure of the graph. One can derive spacetime metrics from a directed acyclic graph of links and check its isotropic property on a large scale. Both approaches are computationally demanding and yet to be explored for high dimensions.

\textbf{Lorentz invariance.} The link between Lorentzian spacetime manifolds and causal sets is known for quite some time. Spacetime events in a Lorentzian spacetime manifold form a causal set~\cite{Sorkin1987}. In the opposite direction, the mainstream causal set theory (CST) has a central conjecture (also called the Hauptvermutung) that a Lorentzian manifold can approximate any causal set with specific properties in a unique way under isometric transformations. There is a proof of this conjecture for some individual cases~\cite{Meyer1982}.

We have a different approach because the proposed theory generates directed acyclic graphs rather than causal sets. These graphs have partial order between pairs of adjacent vertices, so causal sets appear from them via transitivity of partial order on directed paths of links. In addition to this, the vertices of the graph have a measure of proximity that we have defined as the number of links in the shortest chain of links regardless of their direction $l(v,w)$. We also stick to the paradigm of ``geometry = order + number", but we have to define a different measure of volume and distance because the theory does not have the required properties for a ``faithful embedding" of a causal set into a spacetime manifold. Here is a review on the CST for refferences~\cite{Wallden2013}.

\textbf{Spacetime interval.} Now it is time to introduce an approach to define a covariant measure of spacetime interval between any two events. We start with a definition for a spacelike spacetime interval. 

\begin{definition}

Spacelike spacetime interval can be defined for any two spatially separated events $v$ and $w$ that do not have any directed path of links between them and have a joint causal future $J^{+}(v)\cap J^{+}(w)\ne\emptyset$. These events always have a common past $J^{-}(v)\cap J^{-}(w)$ with at least the initial event $b$. A positive integer value of spacelike spacetime interval equals to:

\begin{subequations}\label{eq:interval_spacelike}
\begin{eqnarray}
s^{2}_{s}(v,w)=\min\limits_{u,y}[
&& (|l_{+}(u,w)|+|l_{+}(v,y)|)\times\\
&& (|l_{+}(u,v)|+|l_{+}(w,y)|)],\\ 
s.t.~
&& u\in J^{-}(v)\cap J^{-}(w),\\
&& y\in J^{+}(v)\cap J^{+}(w)
\end{eqnarray}
\end{subequations}

\end{definition}

\begin{figure}
\begin{center}
\includegraphics[width=0.85\linewidth]{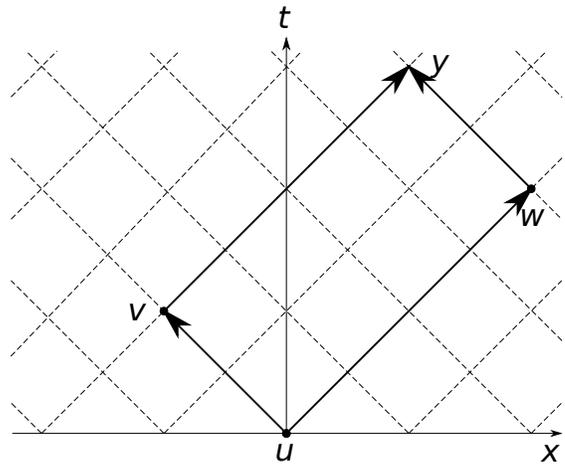}
\end{center}
\caption{A schematic example for calculation of spacelike spacetime interval $s^{2}_{uw}$ between events $u$ and $w$ by finding direct paths $l_{+}(u,v)$, $l_{+}(v,y)$, $l_{+}(u,w)$ and $l_{+}(w,y)$, that minimize the value of a positive spacelike spacetime interval.}
\label{fig:FIG_5}
\end{figure}

We demonstrate the logic behind this definition with a flat spacetime example of a two-dimensional square lattice shown on Figure~\ref{fig:FIG_5}. We assume that all directions of edges of this graph go from bottom to top. In this example, the frame of reference has the t-axis along one diagonal of the square lattice and the x-axis along the other diagonal of the square lattice. The coordinates of all events are the following: $t_u=0$, $x_u=0$, $t_v=2$, $x_v=-2$, $t_w=4$, $x_w=4$, $t_y=6$, $x_y=2$. Then the spacetime interval between events $v$ and $w$ equals to

\begin{subequations}
\begin{eqnarray}
s^{2}_{vw}
&&=(x_w-x_v)^2-(t_w-t_v)^2\label{eq:interval_true}\\
&&=((x_w+t_w)-(x_v+t_v))\\
&&\times((t_v-x_v)+(x_w-t_w))\\
&&=(2 |l_{+}(u,w)|-0)\times(2 |l_{+}(v,y)|+0)\\
&&=(|l_{+}(u,w)|+|l_{+}(v,y)|)\label{eq:interval_def1}\\
&&\times(|l_{+}(u,v)|+|l_{+}(w,y)|)\label{eq:interval_def2}\\ 
&&=8*4=32,
\end{eqnarray}
\end{subequations}

where we used equations $|l_{+}(u,w)|=|l_{+}(v,y)|$ and $|l_{+}(u,v)|=|l_{+}(w,y)|$ for a case of flat spacetime. In a general case of non-flat spacetime we simply postulate Eq.~\ref{eq:interval_def1}-\ref{eq:interval_def2} for Eq.~\ref{eq:interval_true}. 

\begin{definition}

Timelike spacetime interval can be defined for any two timelike separeated events $v$ and $w$ that have a path of links $l_{+}(v,w)$. A negative integer value of timelike spacetime interval equals to:

\begin{subequations}\label{eq:interval_timelike}
\begin{eqnarray}
s^{2}_{t}(v,w)=-\max\limits_{u,y}\left[s^{2}_{s}(u,y)\right],\\ 
s.t.\ u,y\in J^{+}(v)\cap J^{-}(w)
\end{eqnarray}
\end{subequations}

\end{definition}

As a result, we have defined spacetime interval $s^2_{vw}$ for any two events that have shared future purely from the structure of a directed acyclic graph. This measure is quantized, and it is a multiplication of two integers. It is either positive, negative or zero. This definition for spacetime interval is different from that in the causal set theory~\cite{Gregory1991} because we do not postulate a unit volume for each event, but rather provide an explicit formula for spacetime interval.

\textbf{Time.} We define proper time on a world line or any path of links $l_{+}(v,w)$ as an integral sum over square roots of infinitesimal timelike spacetime intervals along the world line. Since the spacetime interval between adjacent events is frequently zero while it is non-zero on a larger scale, we take the maximum value for an integral sum over all segmentations of the world line with the length of each segment not more than a small number $\epsilon$, which represents the size of a clock that measures time.

\begin{definition}

Proper time along a path $l_{+}(v,w)$ is defined by a formula:

\begin{equation}\label{eq:proper_time}
\Delta\tau(l_{+}(v,w))=\max{\sum\limits_{l_{+}(u,y)\in l_{+}(v,w)}\sqrt{-s^2_{uy}}}
\end{equation}

\end{definition}

\begin{definition}

Maximum proper time between any two causally related events with multiple possibilities of world lines between them is defined by:

\begin{equation}\label{eq:max_proper_time}
\tau_{max}(v,w)=\max\limits_{l_{+}(v,w)}{\Delta\tau(l_{+}(v,w))}
\end{equation}

\end{definition}

\textbf{Distance.} We define proper length or rest length along a chain $l(v,w)$ between causally unrelated events $v$ and $w$ in a similar way.

\begin{definition}

Proper length between two causally unrelated events $v$ and $w$ is a minimum sum over square root of infinitesimal spacelike spacetime intervals along the chain $l(v,w)$:

\begin{equation}\label{eq:proper_length}
\Delta L(l(v,w))=\min{\sum\limits_{l(u,y)\in l(v,w)}\sqrt{s^2_{uy}}}
\end{equation}

\end{definition}

Again, this definition is more straightforward than the definition in the causal set theory, and it does not require special consideration for 3+ dimensional spacetime, where a CST definition always gives the value of two as an answer~\cite{Wallden2009}.

\begin{figure}
\begin{center}
\includegraphics[width=1.0\linewidth]{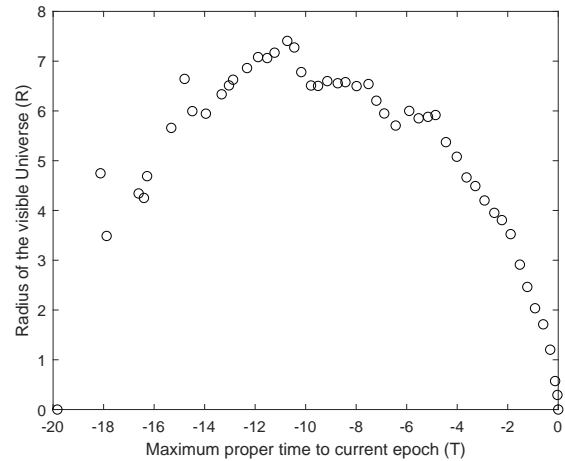}
\end{center}
\caption{Dependence of the radius of the visible Universe $R_{-t}(v)$ versus the maximum proper time $T_{-t}(v)$ to the current epoch for an analog of Model A in four dimensions for 3128 events $v$ at epoch 50 with descendants at epoch 75. This dependence shows the edge of the past light cone that is linear for small times $T_{-t}(v)$ with an angle consistent with the speed of light $c=1$. Every point is an average over all events $v$ for each epoch $t$.}
\label{fig:FIG_6}
\end{figure}

On Figure~\ref{fig:FIG_6} we show the result of calculation for the edges of past light cones in four dimensions. As expected, the growth of the average radius of these past light cones is linear to the maximum proper time with an angle approximately corresponding to the speed of light $c=1$. The average radius also decreases to zero while past light cones approach the initial event $b$.

\textbf{Inertial frame of reference.} We define an inertial frame of reference in the following way. Let us consider a region of spacetime far from the initial event $b$ and away from large-scale curvature anomalies. Suppose there are two events $v$ and $w$ with the shortest path between them and the maximum proper time $\tau_{max}(l_{+}^{*}(v,w))$ that defines a world line for the origin of the frame of reference. We measure time by the proper time along the path $l_{+}^{*}(v,w)$, starting from the event $v$. An event $x$ has the same time as an event $y\in l_{+}^{*}(v,w)$ if the event $y$ maximizes spacelike spacetime interval $s^{2}_{xy}$ between them. The distance between two events with equal times is simply a spatial distance between them. These time and distance define a frame of reference $FR_{1}$. We call it an inertial frame of reference. In general, the lines of equal time and distance coordinates cross at an angle that depends on the choice of the frame of reference, which is set by a pair of events $v$ and $w$. We keep a more strict consideration out of the scope of this paper.

\begin{figure}
\begin{center}
\includegraphics[width=0.85\linewidth]{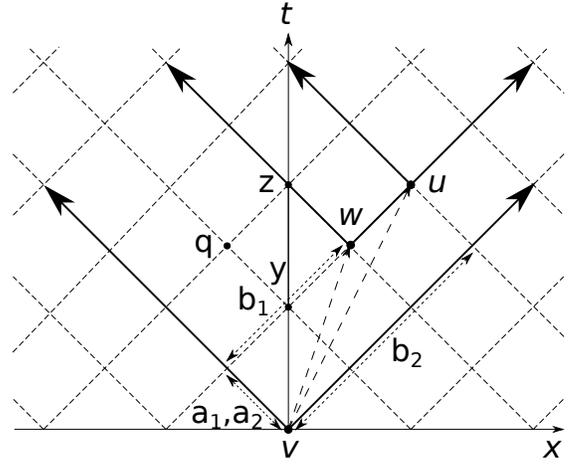}
\end{center}
\caption{An illustration that a moving object $w$ that is emmiting $k$ equidistant light cones at frequency $f_{vw}=k/\tau_{vw}$ is seen with a lower frequency of emission $f_{vu}=k/\tau_{vu}$ by a faster moving object $u$. Events $y$, $z$ and $q$ are introduced for calculation of coordinates of the event $w$.}
\label{fig:FIG_7}
\end{figure}

Stationary points in $FR_{1}$ are spacetime events that have world lines that are parallel to the path $l_{+}^{*}(v,w)$, i.e., for every event $u\in l_{+}^{*}(v,w)$ the maximal spacelike spacetime interval to the events on these parallel world lines is the same up to a small value. On the contrary, a point moving with a constant speed in the frame of reference $FR_{1}$ have a world line that is the shortest path between two events on world lines of different stationary points. Suppose a moving point passes by an event $v$, so its world line is along a path $l_{+}^{*}(v,u)$. The angle between $l_{+}^{*}(v,w)$ and $l_{+}^{*}(v,u)$ defines the speed of the moving point, which is equal to the spatial distance from $u$ to $l_{+}^{*}(v,w)$ divided by the time for the event $u$. See Figure~\ref{fig:FIG_7} for an illustration.

\textbf{Doppler effect.} The initial inspiration for this theory came from its ability to demonstrate mechanisms for various redshifts. We define a redshift (blueshift) as an increase (decrease) of maximum proper time between two points at the edges of two nested light cones.

The simplest example of a redshift or a blueshift is trivial for a change of coordinates from one inertial frame of reference to another, taking into account the definition of spacetime interval. This is a demonstration of the Doppler effect~\cite{Doppler1842}. 

An example of the Doppler effect with a redshift can be demonstrated within the considered frame of reference $FR_{1}$, which is defined by events $v$ and $w$. Suppose there is a clock that moves along the path $l_{+}^{*}(v,w)$ and the maximum proper time between $v$ and $w$ is $\tau_{vw}$. The clock emits information at every tick. Suppose there are $k$ equidistant ticks at events $w_{i}\in l_{+}^{*}(v,w)$, so that $w_{0}=v$ and $w_{k}=w$. The frequency of emission is $f_{vw}=k/\tau_{vw}$. Every tick at event $w_{i}\in l_{+}^{*}(v,w)$ emits a future light cone $J^{+}(w_{i})$, so that the points at the edges of these nested light cones move away from the origin of the frame of reference $FR_{1}$ with a maximal speed of information propagation or the speed of light. If we take any event at the edge of the last light cone $u\in J^{+}(w)$ then we can define another inertial frame of reference $FR_{2}$ with its origin moving along the shortest path $l_{+}^{*}(v,u)$ away from the origin of the frame of reference $FR_{1}$. The maximum proper time between $v$ and $u$ is $\tau_{vu}$. Every edge of a light cone $J^{+}(w_{i})$ crosses the path $l_{+}^{*}(v,u)$, so the tick of the clock have frequency $f_{vu}=k/\tau_{vu}$ in the second frame of reference $FR_{2}$. The redshift happens because $\tau_{vu}\ge\tau_{vw}$.

The proof that $\tau_{vu}\ge\tau_{vw}$ is straigtforward. By definition of spacetime interval for a flat space: $s^{2}_{vw}=-4a_1b_1$, $s^{2}_{vu}=-4a_2b_2$, and $|s^{2}_{vu}|\ge|s^{2}_{vw}|$, because $J^{+}(u)\in J^{+}(w)\in J^{+}(v)$, so that $a_2=a_1$ and $b_2\ge b_1$. Finally, from the definition of proper time: $\tau_{vu}=\sqrt{-s^{2}_{uu}}\ge\sqrt{-s^{2}_{uw}}=\tau_{vw}$. See Figure~\ref{fig:FIG_7} for an illustration.

\textbf{Relativistic sum of velocities.} Derivation of the exact Dopler formula and a relativistic sum of two velocities can be done in a similar way from the definition of spacetime interval, proper time and proper distance. First, we define velocity versus global stationary frame of reference. We calculate time $t_{vw}$ and distance $x_{vw}$ coordinates of the event $w$ versus the event $v$ in this frame of reference from the following logic. The time coordinate of the event $w$ is equal to two terms: $t_{vw}=t_{vy}+t_{yw}$, where we define $t_{vy}=\sqrt{-s^{2}_{vy}}=\sqrt{4a^{2}_{1}}=2a_1$ and measure $t_{yw}$ coordintate along a light like spacitime inteval $s^2_{yw}=0$, so that $t_{yw}=\frac{1}{2}\sqrt{-s^{2}_{yz}}=\frac{1}{2}\sqrt{4(b_1-a_1)^2}=b_1-a_1$. As a result $t_{vw}=b_1+a_1$. The distance coordinate is calculated from the corresponding space like spacetime interval: $x_{vw}=\frac{1}{2}\sqrt{s^{2}_{qw}}=\frac{1}{2}\sqrt{4(b_1-a_1)^2}=b_1-a_1$. Now, $v_{vw}=x_{vw}/t_{vw}$.

Finally, we get a genearal formula of velocity for a world line with a timelike spacetime interval $s^2=4ab$ with corresponding paths $a$ and $b$ along a future light cone from the origin of the frame of reference:

\begin{equation}\label{eq:velocity}
v=\frac{b-a}{b+a}
\end{equation}

So we have two velocities $v_1=(b_1-a_1)/(b_1+a_1)$ and $v_2=(b_2-a_2)/(b_2+a_2)$, and we need to find a formula for velocity $v^{\prime}_{2}$ corresponding to velocity $v_2$ in the frame of reference $FR_{1}$ that is moving with velocity $v_1$. For this purpose, we need to calculate coordinates $t^{\prime}_{vu}$ and $x^{\prime}_{vu}$ in the frame of reference $FR_{1}$. We use the following definition of spacetime interval: $s^{2}_{vu}=-4a_2b_2=x^{2}_{vu}-t^{2}_{vu}=x^{\prime2}_{vu}-t^{\prime2}_{vu}$. And we use the following relation between the coordinates in the frame of reference $FR_{1}$: $t^{\prime}_{vu}=t^{\prime}_{vw}+t^{\prime}_{wu}$ and $x^{\prime}_{vu}=x^{\prime}_{wu}$. Then we use the definition of proper time $t^{\prime}_{vw}=\tau_{vw}=\sqrt{4a_1b_1}$ and the definition of speed of light $x^{\prime}_{wu}=t^{\prime}_{wu}$. Finally, we have an equation:

\begin{equation}\label{eq:xt_prime}
-4a_2b_2=x^{\prime2}_{wu}-(\sqrt{4a_1b_1}+x^{\prime}_{wu})^2,
\end{equation}

which gives a solution to $x^{\prime}_{vu}=(a_2b_2-a_1b_1)/\sqrt{a_1b_1}$ and $t^{\prime}_{vu}=(a_2b_2+a_1b_1)/\sqrt{a_1b_1}$. As a result, we calculate the desired velocity:

\begin{equation}\label{eq:v_prime}
v^{\prime}_{2}=\frac{a_2b_2-a_1b_1}{a_2b_2+a_1b_1}
\end{equation}

And using the fact that $a_1=a_2$ we come to a well known formula of a relativistic sum of velocities:

\begin{equation}\label{eq:doppler_velocities}
\frac{1-v_2}{1+v_2}=\frac{1-v^{\prime}_{2}}{1+v^{\prime}_{2}}\frac{1-v_1}{1+v_1}
\end{equation}

Finally, we get the exact formula for the Doppler effect:

\begin{equation}\label{eq:doppler_frequencies}
\frac{f^{\prime}}{f}=\frac{\tau_{vw}}{\tau_{vu}}=\sqrt{\frac{a_1b_1}{a_2b_2}}=\sqrt{\frac{1-v^{\prime}_{2}}{1+v^{\prime}_{2}}}
\end{equation}

\begin{figure}
\begin{center}
\includegraphics[width=1.0\linewidth]{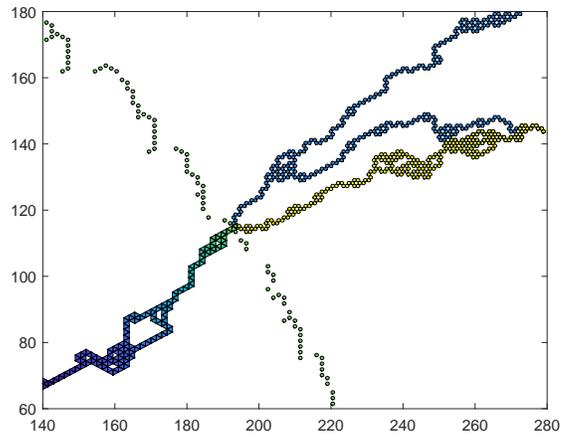}
\end{center}
\caption{A demonstration of strong devergence of future light cones $J^{+}(x)$ and $J^{+}(y)$ (light blue circles and yellow circles) from each other due to local and global curvature of spacetime, regardless the fact that two close events $x$ and $y$ have a common preceeding event $v$. Green circles show an epoch with $n=313$ for the event $v$. Events from interiors of black holes are not shown.}
\label{fig:FIG_8}
\end{figure}

\textbf{Gravitational redshift.} The extremal world lines at the edges of light cones also diverge from each other due to the local curvature of spacetime near black hole singularities. In order to see this effect at the Planck scale, we show an example of two future light cones originating from two close descendants of an event near an event horizon of a black hole, see Figure~\ref{fig:FIG_8}. The edges of these two future light cones diverge quickly from each other as it happens in the real Universe for world lines of two light beams originating from a small region near an event horizon of a black hole.

\textbf{Cosmological redshift.} There is global curvature in the considered theory due to the presence of the initial event $b$ at the center of the graph. This global curvature causes continuous divergence of the edges of two future light cones originating from two close events, see Figure~\ref{fig:FIG_8} for an example. The same logic applies to the past light cones.

Now we define the Hubble constant from the dependance of the size of the visible Universe and the time passage from that epoch~\cite{Hubble1929}. We take an event $v$ at epoch $n$ and all events in its past light cone that belongs to epoch $n-t$: $w\in S_{-t}(v)=J^{-}(v)\cap E(n-t)$. We take the maximum proper time between these events:

\begin{equation}\label{eq:hubble_time}
T_{-t}(v)=\max\limits_{w\in S_{-t}(v)}\tau(l_{+}(w,v)),
\end{equation}

which represents the amount of time that has passed from epoch $n-t$ to the event $v$ along a stationary world line originating from the initial event $b$. We calculate the maximum diameter between the events from the spacetime region $S_{-t}(v)=J^{-}(v)\cap E(n-t)$ as a maximum of minimum proper distance between two events in this region:

\begin{equation}\label{eq:hubble_diameter}
D_{-t}(v)=\max\limits_{u,y\in S_{-t}(v)}\min L(p(u,y))
\end{equation}

We now define a radius of the spacetime region $S_{-t}(v)$ as a maximum distance from the event that maximizes proper time $T_{-t}(v)$. At a large scale this radius is a half of the diameter: $R_{-t}(v)=D_{-t}(v)/2$. In flat space $R_{-t}(v)=T_{-t}(v)$, but in a general case $R_{-t}(v)\ne T_{-t}(v)$ becasue of spacetime curvature. Then we define cosmological redshift from the formula:

\begin{equation}\label{eq:redshift}
1+z_{-t}(v)=\frac{T_{-t}(v)}{R_{-t}(v)}
\end{equation}

The redshift $z$ is zero at $T=0$, and it is positive on a large scale due to positive overall curvature and expansion of space. The linear term in the Taylor expansion of $z(T)$ is related to the Hubble constant, while the quadratic term is additionally influenced by the local curvature: $z_{-t}(v)=f(T_{-t}(v))\approx H(v)T_{-t}(v)$, where $H(v)$ is an approximate value of the Hubble constant around the event $v$. It is measured in Planck units. Finally, the value of the Hubble constant at epoch $n$ is an average for each event from this epoch: $H(n)=\overline{H(v)}$.

Calculation of the Hubble constant from the ratio $z_{-t}(v)/T_{-t}(v)$ at $t\approx 20$ for an analog of Model A in three dimensions gives an approximate value of $H(n)\sim\frac{1}{n}$, which is in line with the current estimation for $H$.

\textbf{Metric tensor.} Now we introduce an embedding of a directed acyclic graph $\mathcal{G}$ into a pseudo-Riemannian manifold $\mathcal{M}$ with a mapping $\eta$ that:
\begin{itemize}
  \item presersves causality: $\exists l_{+}(v,w)\Leftrightarrow\eta(v)\prec\eta(w) $
  \item preserves spacetime interval values $s^2(v,w)$ for any events $v$ and $w$ with common future
\end{itemize}

There is an open question whether this mapping $\eta$ exists. The graph $\mathcal{G}$ is a lattice, so there is no problem with embedding it into a Euclidian manifold. The critical question is whether a pseudo-Riemannian signature metric tensor $g_{\mu\nu}$ exists that is consistent with all causal relations and spacetime intervals.

The mapping $\eta$ is also non-unique at least for small deformations of $\mathcal{M}$ at the scale smaller than a link of the graph. There is a proven theorem that a Lorentzian geometry $\mathcal{M}$ follow from causal order only up to a local conformal factor ~\cite{Robb1914},~\cite{Robb1936},~\cite{Reichenbach1921},~\cite{Zeeman1964}. Unit volume $\sqrt{-g}dx^4$ fixes this local conformal factor in the mainstream causal set theory. The CDFT, on the contrary, fixes timelike and spacelike spacetime intervals between the vertices of a directed acyclic graph.

In order to move forward, we formulate a variant of the central conjecture for the causal discrete field theory:

\begin{quote}
Every directed acyclic graph $\mathcal{G}$ from the causal discrete field theory can be uniquely embedded into a pseudo-Riemannian manifold up to an isometric transformation: $\mathcal{M}_1(\mathcal{G})\approx\mathcal{M}_2(\mathcal{G})$
\end{quote}

We propose an approximate way to calculate the metric tensor $g_{\mu\nu}(v)$ for each event $v$ from the geometry of its vicinity $V(v,\epsilon)$ that is $\epsilon$ links apart by minimizing the fitting error of the local metric tensor to reproduce local spacetime intervals.

First, we assume some vector coordinates for an n-dimensional case with the center at the initial vertex $b$. We use unit length basis vectors along $n$ chosen graph links originating from the initial vertex $b$. Any choice of $n$ links out of $n+1$ is possible because the graph is self-similar. The coordinates of any event appear from parallel translations of this unit length basis vectors. This definition of coordinates is proper for simple self-similar graph geometries like cubic lattice or generalized diamond lattice. Some exotic self-similar graphs may need a different approach with non-unique coordinates for the same events, e.g., for a finite graph like a fullerene molecule or a graph with some compactified finite dimensions like a nanotube molecule.

Second, we take all events from the vicinity of the givent event: $w\in V(v,\epsilon)$, and calculate coordinate differentials between the events $dx^{\mu}_{vw}=x^{\mu}_{w}-x^{\mu}_{v}$ and actual spacetime intervals between them $s^{2}_{vw}$.

Finally, we minimize square errors between a bi-linear combination of coordinate differentials and actual spacetime intervals:

\begin{subequations}\label{eq:g_mu_nu_epsilon}
\begin{eqnarray}
s^{2}_{vw}=g_{\mu\nu}(v,\epsilon)dx^{\mu}_{vw}dx^{\nu}_{vw}+e_{vw},\\ 
s.t.\ \sum\limits_{w\in V(v,\epsilon)}{e^{2}_{vw}} \rightarrow min
\end{eqnarray}
\end{subequations}

As a result, we calculate an $\epsilon$-averaged metric tensor $g_{\mu\nu}$, where $\epsilon$ is the size of the averaging region measured in the number of links from the chosen event $v$.

\textbf{Spacetime curvature.} We propose to do an approximate calculation of all relevant elements of Riemannian geometry. We can take approximate values of metric tensor at each event $g_{\mu\nu}(v,\epsilon)$ and calculate a twice differentiable interpolation of it beyond the vertices of the graph. It allows to calculate partial derivatives of the metric tensor:

\begin{equation}\label{eq:g_derivatives}
\frac{\partial}{\partial x^k}g_{ij}=g_{ij,k}\ \ \text{and}\ \  \frac{\partial^2}{\partial x^k\partial x^l}g_{ij}=g_{ij,kl} 
\end{equation}

And then combine them into the Christoffel symbols:

\begin{subequations}\label{eq:christoffel_definition}
\begin{eqnarray}
\Gamma_{kij} && =\frac{1}{2}\left(\frac{\partial}{\partial x^j}g_{ki}+\frac{\partial}{\partial x^i}g_{kj}-\frac{\partial}{\partial x^k}g_{ij}\right)\\ 
&& =\frac{1}{2}(g_{ki,j}+g_{kj,i}-g_{ij,k})
\end{eqnarray}
\end{subequations}

\begin{equation}\label{eq:christoffel}
\Gamma^{m}_{\ ij}=g^{mk}\Gamma_{kij}=\frac{1}{2}g^{mk}(g_{ki,j}+g_{kj,i}-g_{ij,k})
\end{equation}

Finally, the Riemannian curvature tensor equals to

\begin{equation}\label{eq:curvature_tensor_definition}
R^{l}_{\ ijk}=\frac{\partial}{\partial x^{j}}\Gamma^{l}_{\ ik}-\frac{\partial}{\partial x^{k}}\Gamma^{l}_{\ ij}+\Gamma^{l}_{\ js}\Gamma^{s}_{\ ik}-\Gamma^{l}_{\ ks}\Gamma^{s}_{\ ij}
\end{equation}

Lowering the indexes $R_{sijk}=g_{sl}R^{l}_{\ ijk}$ we get

\begin{subequations}\label{eq:curvature_tensor}
\begin{eqnarray}
R_{sijk} = && \frac{1}{2}(g_{sk,ij}+g_{ij,sk}-g_{sj,ik}-g_{ik,sj})+\\ 
&& +g_{np}(\Gamma^{n}_{\ ij}\Gamma^{p}_{\ sk}-\Gamma^{n}_{\ ik}\Gamma^{p}_{\ sj})
\end{eqnarray}
\end{subequations}

Finally, Ricci curvature tensor and the Ricci curvature equal to

\begin{equation}\label{eq:ricci}
R_{ij}=g^{lm}R_{iljm}\ \ \text{and}\ \ \mathcal{R}=g^{ij}R_{ij}
\end{equation}

We have described a way to calculate approximate metric tensor and approximate curvature tensor appealing to continuous Riemannian geometry. It is helpful, for example, to calculate large-scale curvature of the resulting spacetime. However, it is still interesting to find a way to define discrete analogs of these objects purely from the structure of a directed acyclic graph. 

\textbf{Energy and momentum.} Now we introduce a trivial definition of energy and momentum. There is already a metric tensor $g_{ij}$ in this theory. This tensor field defines all objects within Riemannian geometry on a manifold, including the Einstein tensor:

\begin{equation}\label{eq:einstein_tensor}
G_{ij}=R_{ij}-\frac{1}{2}\mathcal{R} g^{ij}
\end{equation}

The Einstein tensor is symmetric $G_{ij}=G_{ji}$ and it is divergenceless $\nabla_{i}G^{ij}=G^{ij}_{;i}=0$. The stress-energy tensor of matter and the metric tensor have the same property, so we can postulate the stress-energy tensor to be equal to the Einstein tensor up to a cosmological term given the Planck units assumption of $G=1$ and $c=1$:

\begin{equation}\label{eq:stress_energy_tensor}
T_{ij}=\frac{1}{8\pi}\left(G_{ij}+\Lambda g_{ij}\right)
\end{equation}

A more strict approach requires an understanding of the local dynamics of discrete fields on the graph. We use the Einstein field equations to reconstruct the energy-momentum tensor because gravitational field emerges selfconsistently without equations of matter. We develop the theory from pure geometry of the graph and the propagation rule.

The role of the cosmological term $\Lambda$ is still not clear. This term can also be variable, which leads to the transfer of energy and momentum between the cosmological term and matter. The implications of a variable cosmological term for evolution of the scale factor $a(t)$ was studied in literature with different assumptions for decay laws~\cite{Overduin1998}. Our favorable hypothesis is to use it to offset the global curvature of spacetime. The idea is to have an energy-momentum tensor $T_{ij}$ that is only due to the local dynamics of fields on the graph. As a result, the cosmological term would be positive and proportional to the inverse square of epoch number $n$: $\Lambda = \mathcal{R}(n)\sim\frac{1}{n^2}$, which is in line with the current estimation for $\Lambda$.

\begin{figure}
\begin{center}
\includegraphics[width=1.0\linewidth]{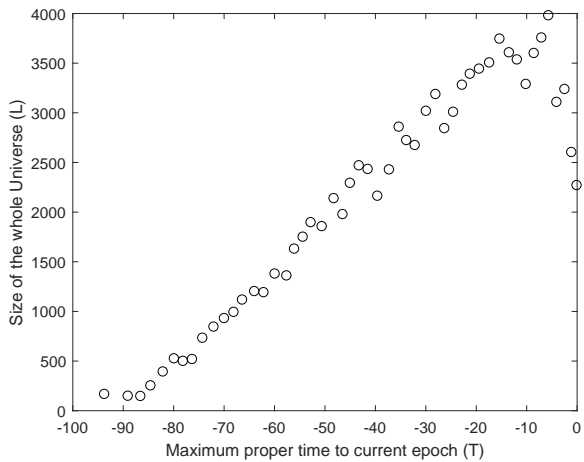}
\end{center}
\caption{Dependence of the size of the whole Universe $L_{-t}(v)$ versus the maximum proper time $T_{-t}(v)$ to the current epoch for an analog of Model A in three dimensions for 1747 events $v$ at epoch 200 with descendants at epoch 250. Every point is an average over all events $v$ for each epoch $t$.}
\label{fig:FIG_9_3D}
\end{figure}

\begin{figure}
\begin{center}
\includegraphics[width=1.0\linewidth]{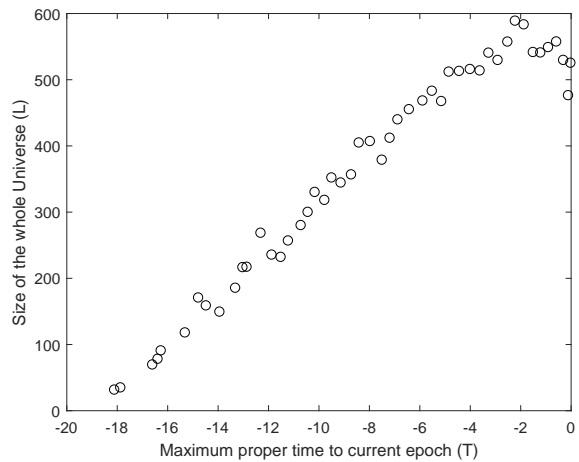}
\end{center}
\caption{Dependence of the size of the whole Universe $L_{-t}(v)$ versus the maximum proper time $T_{-t}(v)$ to the current epoch for an analog of Model A in four dimensions for 3128 events $v$ at epoch 50 with descendants at epoch 75. Every point is an average over all events $v$ for each epoch $t$.}
\label{fig:FIG_9_4D}
\end{figure}

\textbf{Inflation.} Computable Model A does not reproduce the inflation scenario right after the Big Bang. It is an open question to check this matter for non-computable models. The inflationary expansion of the Universe~\cite{Guth1981} was proposed to explain the uniformity of cosmic background radiation~\cite{Wilson1965} that has not interacted with matter since the moment of the photon decoupling event~\cite{Zeldovich1968}.

We calculate all spacetime intervals and proper times for an analog of Model A in three and four dimensions because the two-dimensional case is degenerate. We estimate the size of the whole Universe at each epoch $n-t$ from the radius of the visible Universe $R_{-t}(v)$ and its angular size $\phi_{-t}(v)$ in the embedding space of the graph:

\begin{equation}\label{eq:Universe_size}
L_{-t}(v)=\frac{2\pi}{\phi_{-t}(v)}R_{-t}(v)
\end{equation}

The size of the Universe $L_{-t}(v)$ depends linearly from the proper time $T_{-t}(v)$, see Figures~\ref{fig:FIG_9_3D} and \ref{fig:FIG_9_4D}. In four dimensions: $T\approx -0.39 t$ and $L \approx 2\pi\times 2.1 (n-t)$.

We also compare our numeric calculations with a continuous limit. In this case the edge of the past light cone is described by an equation in polar coordinates of the angle $\phi$ and the overall epoch number $t$:

\begin{equation}\label{eq:epoch_differential}
d\phi=\frac{c\alpha dt}{t}
\end{equation}

where we have introduced a rate of time flow which corresponds to an angle of the edge of the past light cone in the embedding space of the underlying graph: $\alpha=\frac{2\pi t}{L}\frac{d\tau}{dt}$. With a constant speed of light $c=1$ and a constant rate of time flow $\alpha$, the angle coordinate for the edge of the past light cone starging at epoch $t_0$ will depend on the past epoch $t_1$ in the following way:

\begin{equation}\label{eq:epoch_integral}
\Delta\phi=\alpha\ln(\frac{t_0}{t_1})
\end{equation}

This formula shows that if one looks back for long enough, then the whole Universe can be seen an infinite number of times in the continuous limit. If in practice the backward propagation stays within a limited angle because the rate of time flow $\alpha$ is much lower for smaller epochs, then a distant observer at the current epoch $t_0$ will perceive the Universe expanding exponentially fast close to the Big Bang event.

\begin{figure}
\begin{center}
\includegraphics[width=1.0\linewidth]{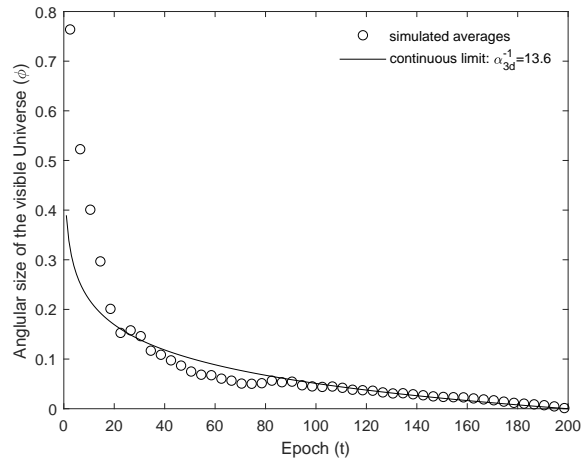}
\end{center}
\caption{Dependence of the angular size of the visible Universe $\phi_{-t}(v)$ from epochs $t$: simulated points (circles) and continuous limit with $\alpha_{3d}\approx \frac{1}{13.6}$ (solid line). The calculation is done for an analog of Model A in three dimensions for 1747 events $v$ at epoch 200 with descendants at epoch 250. Every point is an average over all events $v$ for each epoch $t$.}
\label{fig:FIG_10_3D}
\end{figure}

\begin{figure}
\begin{center}
\includegraphics[width=1.0\linewidth]{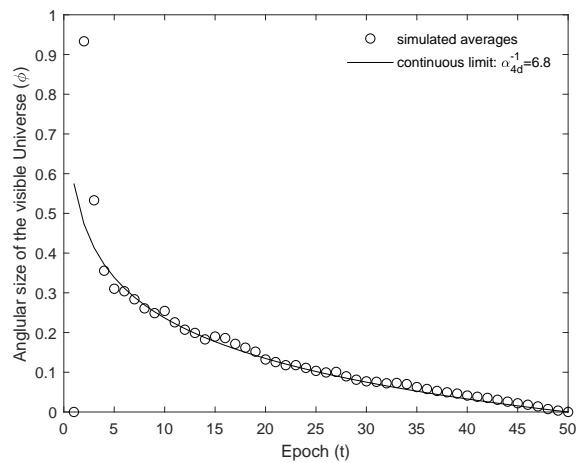}
\end{center}
\caption{Dependence of the angular size of the visible Universe $\phi_{-t}(v)$ from epochs $t$: simulated points (circles) and continuous limit with $\alpha_{4d}\approx \frac{1}{6.8}$ (solid line). The calculation is done for an analog of Model A in four dimensions for 3128 events $v$ at epoch 50 with descendants at epoch 75. Every point is an average over all events $v$ for each epoch $t$.}
\label{fig:FIG_10_4D}
\end{figure}

We show our calculations for the angular size of the visible Universe on Figures~\ref{fig:FIG_10_3D} and \ref{fig:FIG_10_4D}. We make the best fit for $\alpha_{3d}\approx\frac{1}{13.6}$ in three dimensions and $\alpha_{4d}\approx\frac{1}{6.8}$ in four dimensions, which represents the angle for the edges of past light cones at large epochs and approximately equals to $\alpha\approx\frac{2\pi t}{L}\times\frac{d\tau}{dt}$. It would be interesting to find the exact value of $\alpha$ for Model A with large $n$.

The simulated dependence in three dimensions is close to the continuous limit with slightly slower expansion during the early epochs and marginally faster expansion at the middle of the plot. The simulated dependence in four dimension fits the line of the continuous limit even better, but it has only 50 calculated epochs. This computational result suggests that Model A generates a simple cosmological model~\cite{Nanopoulos1996} with a variable cosmological term $\Lambda\sim\frac{1}{\tau^2}\sim\frac{1}{L^2}$, which avoids the problem of small fundamental constants~\cite{Dirac1937},~\cite{Dirac1938}. The resulting metrics depends on the epoch $t$ similar to the Milne model~\cite{Milne1935} but with positive curvature:

\begin{equation}\label{eq:metrics}
ds^2=-c^2dt^2+\frac{t^2}{\alpha^2}\left(\frac{dr^2}{1-r^2}+r^{2}d\Omega^2\right)
\end{equation}

The inflationary scenario requires that simulated angles for early epochs are located much lower than the theoretical line for the continuous limit. An alternative explanation for a homogeneous Universe at the photon decoupling event is that the Universe was causally connected at early epochs. In the continuous limit of four dimensions, the edges of the past light cone can do three circles around the Big Bang event during $t_{0}\approx 8\times 10^{60}$ Planck units of time, and the visible Universe coincided with the whole Universe at the moment of $\Delta\phi=\pi$ at the age of the Universe of $t_{1}\approx 7.3$ years.

\textbf{CPT invariance.} Symmetry plays an essential role in theoretical physics. So far, we have discussed translational symmetry, which follows from the fact that each vertex has the same local dynamics due to a propagation rule. We also conjectured that rotational symmetry and isotropy of space follow from isotropic propagation rules for some types of graphs, e.g., for generalized diamond crystal lattices. Now we turn to another fundamental symmetry for matter and anti-matter, which is described by CPT invariance.

Time reversal symmetry changes the direction of all directed links on the graph: $\psi_{ij}=+c\Leftrightarrow\psi_{ij}=-c$. Parity symmetry makes a mirror-image of the graph. Charge symmetry permutes integer values on edges: $i^{\prime}=C(i)$, where $C(i)$ is an integer defined function that exchanges some pairs of integers or keeps some integers unchanged.

Each of these symmetry transformations matches the graph with itself and changes the pattern of discrete fields on it. We call two propagation rules symmetrical if and only if for each pattern on a graph generated by one propagation rule, there is a symmetric pattern on a graph generated by the other propagation rule. C, P or T symmetric propagation rules can appear from the initial propagation rule by applying $C(i)$ to integers on incoming and outgoing edges, applying a non-parity permutation of edges or by changing the direction of all incoming and outgoing edges.

Each of the three symmetries allows for a non-trivial change of a propagation rule so that the new rule is substantially different from the initial one. A notable case is when the application of all three symmetries simultaneously creates the same propagation rule. We call this type of propagation rules as CPT invariant propagation rules.

We call a propagation rule violating symmetry if and only if the application of this symmetry to all its patterns on graphs substantially changes the resulting propagation rule.

Model A and Model A-D on Figure~\ref{fig:FIG_1} are examples with CPT invariant non-trivial propagation rules. Model A-D also violates C, P, and T symmetries separately, because all four vertices in sector B do not have symmetrical counterparts for all three simple symmetries.

The most exciting types of propagation rules are non-trivial non-computable CPT invariant isotropic propagation rules with violation of separate C, P or T symmetries because they are most similar to what happens in the real Universe. However, these propagation rules are hard to compute, which complicates the analysis.

An interesting topic is the CPT theorem stating that any Lorentz invariant local quantum field theory with a Hermitian Hamiltonian must have a CPT symmetry~\cite{Schwinger1951}. If there is an analog of this theorem for the proposed theory, then CPT invariant propagation rules must be the key for finding a quantum mechanical description for CDFT models.

\textbf{Elementary particles.} There are still no particles in the overall picture. The proposed theory is self-consistent so that it does not require special consideration of matter or elementary particles. However, propagation rules may have one more symmetry of charge conservation, which together with CPT invariance may be important for the introduction of elementary particles in this theory.

Model A-D on Figure~\ref{fig:FIG_1} is a simple example of charge conservation: graph edges with integer $2$ are currents with positive charge and graph edges with integer $3$ are currents with negative charge, and the propagation rule allows only uninterrupted flow of these charges in addition to creation and annihilation of pairs of currents with opposite charges.

Some propagation rules can also have conservation of multiple types of charges, for example, by having multiple pairs of currents like in Model A-D. Also, partial charge conservation is possible, like in the real Universe. For example, there can be strict conservation of ``electric" charge when the number of odd/even integer value incoming edges equal to the number of odd/even integer value outgoing edges for each vertex except for ``neutral" currents like integer $1$ in Model A-D. At the same time, there can be pairs of odd/even integer values $j\Leftrightarrow C(i)$ that are are ``mixed" without parity violation for some rear types of vertices.

\textbf{Mass and inertia.} We now suggest a mechanism of how masses might appear in CDFT models. We already postulated that every field configuration can happen with a likelihood that is defined by a constraint functional $C_{\mathcal{P},b}[\psi]$ in Eq.~\ref{eq:path_integral}. In quantum physics, the principle of least action defines which field configurations or paths of particles are more likely than others. We also follow that approach.

The likelihood of taking an infinite size pattern from an infinite uncountable set is poorly defined, so we start from the likelihood of local configurations for valid types of vertices. We postulate the following ``principle of least action":

\begin{quote}
Sequential growth of vertices is equally likely in terms of the choice of all possible locations and types for the next new vertex.
\end{quote}

This principle makes it easy to construct patterns on self-similar graphs with computable propagation rules by randomly choosing which vertex of which valid type to add next to the growing set of already enumerated vertices. Non-computable propagation rules additionally need to have a non-computable ``oracle" that knows which types of vertices are valid locally given global constraints in the future.

The physical meaning of this principle is the following. A local observer does not care about the global likelihood of future patterns. All that she cares is that there are some possible splits of her Universe on the next vertex of her world line, and she can find herself in any one of them with equal likelihood. This postselection probability calculation resembles the famous sleeping beauty paradox when an observer assigns equal subjective probabilities for indistinguishable future outcomes.

The proposed postselection of Universe splits produces a tremendous difference in the likelihood of valid patterns like it is for possible virtual paths of elementary particles in quantum field theories. Vertices with ``charged" edges are expected to have a lower likelihood on average because charge conservation is an additional constraint that needs to be satisfied. So sequential growth might be slower statistically near conserving charges, which should lead to the emergence of persistent craters with positive Riemannian curvature around them. That is how particle and antiparticle charges might acquire equal positive masses. Gravity might be a result of a massive particle being placed in curved spacetime around another massive particle.

Another direction to explore is whether the sequential growth of vertices allows for charges moving with a constant speed versus the global frame of reference. It is natural to expect stable symmetric craters around charged world lines going perpendicular to the surface of sequential growth. The question is whether locally equiprobable sequential growth can generate stable skewed craters around charged world lines going at an angle to the surface of sequential growth. This would mean that free charges have not only mass but also inertia.

Both ideas for mass and inertia of free moving charges are entirely speculative and up to future confirmation because charge conservation implies non-computable propagation rules that are hard to model computationally. The theory allows for a huge number of field configurations with the possibility to reproduce any complex interaction of particles and spacetime metrics around them, so the question is whether physically meaningful field configurations are more likely due to equiprobable sequential propagation.

\textbf{Entropy.} Locally CPT invariant patterns on a graph still have an essential global asymmetry versus time reversal, because we require only one initial vertex and equal likelihood for local types of vertices during sequential growth. On the contrary, there are a lot of terminal vertices, which are singularity points of black holes. So time reversal of this kind of Universe will lead to only one giant black hole and an infinite number of ``white holes", which are CPT analogs of black holes. In the real Universe, we do have signs of existence for black holes~\cite{EHT2019}, while ``white holes" remain only a theoretical object.

This global time asymmetry may explain the arrow of time and the second law of thermodynamics that forbids the decrease of entropy in a closed system. In the proposed theory the time-reversed Universe indeed collapses to a single state of the terminal point of a giant black hole, so the entropy of the whole Universe eventually drops to the least possible value of zero at the Big Bang event.

Here is an example of a thought experiment to calculate entropy within the proposed theory. We consider a case of a ``stationary" Universe in two dimensions for simplicity. We use a nanotube molecule type of graph with one class of the edges of the graph going along the nanotube. This graph has one infinite dimension along the nanotube and one confined dimension perpendicular to it. The diameter of the nanotube should be large enough to dismiss the effects of circular boundary conditions. Suppose we have $n$ vertices in each circle of the nanotube that go with a period of four given the number of different vertices of a hexagon along the nanotube.

We propose to seed initial vertices on the graph as a circle chain of valid types of vertices around the nanotube. Then we propose to run sequential growth in one of the directions along the nanotube from the initial vertices for long enough to get statistical equilibrium. The direction of propagation will correspond to a global time arrow in this stationary Universe.

We propose to use a CPT invariant charge conserving propagation rule so that the flow of charge through the nanotube is constant given the convention that flow of charge $c$ in the reverse time direction corresponds to a flow of charge $C(c)$ in the global time direction.

\begin{definition}

A microstate $|\Phi_{i}\rangle$ is a set of open-ended graph edges during sequential growth without standalone cavities that eventually collapse into black holes.

\end{definition}

Each step of sequential growth adds one vertex by connecting it to one or more existing open-ended graph edges and forming zero or more new open-ended edges from the new vertex. We exclude cavities from consideration because they are not causally related to the remaining part of the graph any more, and therefore they should be treated as separate microstates.

By running sequential growth, we may reach a statistical equilibrium and calculate the probability for each microstate $|\Phi_{i}\rangle$. The number of all possible microstates is infinite because there is an infinite number of elongated ellipses around the nanotube. However, we expect that circular shape microstates should dominate the probabilities so that infinite sums over all microstates will converge.

\begin{definition}

A macrostate $|\Psi\rangle$ is a set of microstates $|\Phi_{i}\rangle$ that have the same property or symmetry.

\end{definition}

We consider two variants of macrostates in our example. First macrostate has a zero flow of charge: $|\Psi_{0}\rangle$. Second macrostate has a unit flow of charge: $|\Psi_{1}\rangle$. Since we assume a charge-preserving propagation rule, all microstates will stay within their corresponding macrostate during propagation.

Now we can calculate entropy for each macrostate with charge $q$ in equilibrium with a Planck value for the Bolzmann constant $k_{B}=1$:

\begin{equation}\label{eq:entropy}
S_{|\Psi_{q}\rangle}=-\sum\limits_{|\Phi_{i}\rangle\in|\Psi_{q}\rangle}{p_i\ln{p_i}}
\end{equation}

We anticipate that the marcostate with charge $|\Psi_{1}\rangle$ has higher entropy than the macrostate without charge $|\Psi_{0}\rangle$, because due to charge conservation constraints fewer microstates $|\Phi_{i}\rangle$ will have high enough propabilities $p_{i}$.

The delta between the two entropies corresponds to the entropy of a massive particle versus a vacuum state:

\begin{equation}\label{eq:delta_entropy}
\Delta S=S_{|\Psi_{1}\rangle}-S_{|\Psi_{0}\rangle}>0    
\end{equation}

This entropy $\Delta S$ adds another way in addition to the Einstein field equations in Eq.~\ref{eq:stress_energy_tensor} to estimate mass by analogy with the Bekenstein-Hawking entropy of a symmetric Schwarzschild black hole, which is proportional to the square of its mass in four-dimensional spacetime:

\begin{equation}\label{eq:BH_entropy}
S_{BH}=4\pi M^2, 
\end{equation}
where $M$ is the mass of the black hole in Planck units of mass $m_{P}=\frac{\hbar c}{G}=1$.

It is an interesting open question to prove the power law dependence of entropy in Eq.~\ref{eq:delta_entropy} from geometrically defined mass in Eq.~\ref{eq:stress_energy_tensor} in various dimensions and to find the exact coefficient in the equation similar to Eq.~\ref{eq:BH_entropy}.

\textbf{Quantum field theory.} The most challenging task for the CDFT is to explain quantum effects within a causal and discrete paradigm. One alternative is to define a continuous field theory that has a field $\theta$ on some pseudo-Riemannian n-dimensional manifold and an action term $e^{i S[\theta]}$, so that this field theory allows for a quantization constraint that makes a correspondence between the fields $[\theta]\Leftrightarrow[\psi]$ and the action terms $e^{i S[\theta]}\Leftrightarrow C_{\mathcal{P},b}[\psi]$ for a certain self-similar graph.

So far, we have only been able to outline a way to find a correspondence between a discrete field of directions of links $[\psi]$ and a continuous tensor field for metrics $g$ without even construction of the action for the gravity part of the theory.

We propose to search for a variant of a CDFT model that allows for a continuous quantum field approximation that is consistent with the Standard Model of elementary particles at large scale and low-energies in flat spacetime far away from the Big Bang event.

Some progress has been made recently with another approach that shows how a deterministic cellular automata model can demonstrate quantum field theory phenomena. Gerard `t Hooft has studied a simple cellular automata model that is equivalent to string theory in 1+1 dimensions~\cite{tHooft2012}. However, a cellular automata model defines propagation of information on separate slices of time and therefore has a problem describing covariant objects in curved spacetime.

As the next step for future research, we propose to develop these ideas further adding the many-worlds interpretation as an essential ingredient, where a local pattern around one event unitarily evolves to a mix of local patterns around another event.

\textbf{Many-worlds interpretation.} The CDFT is consistent with the many-worlds interpretation of quantum mechanics that took its roots from Erwin Schr$\ddot{o}$dinger's ideas~\cite{Schrodinger1952}, was first described by Huge Everett~\cite{Everett1956},~\cite{Everett1957} and was later popularized by Bryce DeWitt~\cite{DeWitt1973}. The Multiverse in the CDFT contains all copies of Universes, each pair of which has some maximal common area around the Big Bang event. The Multiverse splits any time when the propagation rule allows for more than one configuration of integer fields and edge directions for the same variant of the past. This ambiguity can happen either due to multiple valid types for the next vertex in sequential growth or due to the application of a propagation rule in a different order for this vertex and its direct neighbors.

The CDFT has the following variant of the many-worlds interpretation of quantum mechanics:
\begin{itemize}
  \item Infinite uncountable number of Universes
  \item Finite number of Universes of a given size
  \item Splitting happens only in causal future
  \item No interaction between different Universes
  \item The whole Multiverse is generated by a deterministic propagation rule
  \item The Multiverse is computable for computable propagation rules
  \item The Multiverse may be hard to compute for non-computable propagation rules
\end{itemize}

The resulting Multiverse does not only have multiple variants of Universes, but it also allows for all four non-exclusive possibilities that are considered in literature~\cite{Weinberg2005}. First, sub-universes may be different regions in space. Indeed, some events may not have a shared future, and they may have a shared past only very close to the Big Bang event. Second, fundamental properties may change with time. We postulated that the speed of the fastest information propagation is constant, and we derived time and spatial distance from that. However, some other physical constants may change slowly, for example, the cosmological term $\Lambda$ and the Hubble constant $H$ depend on the global curvature of spacetime. Third, the Universe can have spacetime regions with different properties. Information patterns on a graph may undergo phase transitions while the global curvature changes so that the propagation of information patterns on the graph happens in different ways. Finally, the real Universe can be a subspace of a higher dimensional CDFT model with some dimensionality reduction to a four-dimensional spacetime. 

Apart from quantum mechanics, the discrete version of the many-worlds interpretation provides an interesting view on the von Neumann-Morgenstern probabilities in decision theory~\cite{Neumann1953} and the concept of Turing machines approximating the Universe within the universal artificial intelligence model of Marcus Hutter~\cite{Hutter2005}.

We now briefly discuss some of the quantum phenomena which might be explained by the CDFT in line with a deterministic many-worlds interpretation of quantum mechanics, which has already been discussed in literature~\cite{Vaidman2014}. Moreover, we propose an approach to demonstrate unitary evolution for quantum fields.

\textbf{Heisenberg uncertainty principle.} There is a fundamental limit in quantum mechanics to determine both position and momentum of a particle at the same time~\cite{Heisenberg1927}. Werner Heisenberg first introduced this principle and then Earle Kennard proved it in a general form of $\sigma_{x}\sigma_{p}\ge\frac{\hbar}{2}$, where $\hbar$ is the Planck constant and $\sigma_{x}$ and $\sigma_{p}$ are standard deviations of position and momentum of a particle~\cite{Kennard1927} .

Howard Robertson proved a more general uncertainty relation~\cite{Robertson1929} for any two quantum variables $A$ and $B$: $\sigma_{A}\sigma_{B}\ge\frac{1}{2}\large|\langle\psi|[\hat{A},\hat{B}]|\psi\rangle\large|$. This relation shows that there is a fundamental uncertainty in a simultaneous measurement of two non-commuting quantum variables. This fact is as a manifestation of incomplete description of the Universe by quantum mechanics.

The many-worlds interpretation view on the uncertainty relations is that every quantum measurement splits the Universe so that observers calculate statistical correlations of two consecutive measurements in different splits of the Universe. The simultaneous measurement is only possible for commuting quantum operators because they have common eigenvectors so that the result of the measurement for both operators happens in the same split of the Multiverse.

It is an interesting direction for future research for the CDFT to find an analog of uncertainty relations and to check them computationally.

\textbf{Double-slit experiment.} Classical waves can demonstrate diffraction and interference, but quantum mechanics has a unique twist on top of that. As Richard Feynman noted, the double-slit experiment demonstrates the core phenomenon of quantum mechanics, which cannot be explained in any classical way~\cite{Feynman1965}. In this experiment, the interference is destroyed any time an experimenter measures which path is taken by a particle. David Deutsch argued that this phenomenon is a manifestation of the many-worlds interpretation of quantum mechanics~\cite{Deutsch1998}.

We follow the same explanation. In the CDFT, a quantum measurement happens anytime when the Multiverse splits irreversibly. We conjecture that interference happens each time when multiple Universes coincide everywhere except for a finite causal diamond region of $D(v,w)=J^{+}(v)\cap J^{-}(w)$, so that there are multiple possible field configurations inside $D(v,w)$, so that the future and the past are the same.

On the contrary, the interference disappears whenever splits of the Multiverse began to differ forever after some point along the world line of an observer. The role of the observer is passive. She notices post-factum, which particular variant of the Universe has happened.

\textbf{EPR paradox} is a thought experiment that first demonstrated that quantum mechanics is not a complete description of physical reality~\cite{EPR}. In this experiment, a measurement of a spin for one of two spatially separated spin-singlet particles instantaneously defines the spin for the other particle. Either a ``spooky action at a distance" can describe this, which is contrary to general relativity, or some hidden variables define full deterministic evolution of each particle before the measurement.

The proposed theory has to preserve perfect correlations between the fields along the world lines of particles of the EPR pair in order to explain the paradox in a local and deterministic way. This requirement is quite strong, but still possible in principle. There must be a split of the Multiverse at the spacetime event when the two particles separated from each other. The two variants of the Universe must have identical world lines of two particles except for their spin variables until the next split of the Multiverse when their spin projections are measured.

\textbf{Bell theorem} provides another challenge for a local and deterministic theory. The original Bell's work stated that any local hidden variable classical model could not describe all phenomena in quantum mechanics from an example of two spin-$\frac{1}{2}$ particles forming an EPR pair~\cite{Bell1964}. However, later it was noted that a classical super-deterministic model could allow for satisfaction of Bell-type experiments~\cite{Bell1985}. In this hypothetical model, an observer itself is a part of a fully deterministic environment, so her decisions to choose a polarization of a measurement device depends on the state of the Universe. There is no free will for an observer in such interpretation of the paradox.

A three particle GHZ experiment provides a more clear explanation of the Bell theorem without statistical inequality relations in the original work. It can be shown that a quantum state of three spin-$\frac{1}{2}$ particles $|GHZ\rangle=\frac{1}{\sqrt{2}}(|\uparrow_{z}\rangle_{A}|\uparrow_{z}\rangle_{B}|\uparrow_{z}\rangle_{C}-|\downarrow_{z}\rangle_{A}|\downarrow_{z}\rangle_{B}|\downarrow_{z}\rangle_{C}) \nonumber$ cannot have definite values for x- and y-projections of all spins simultaneously, which means there is no Universe where hidden variables define these projections.

In Lev Vaidman's interpretation of this experiment~\cite{Vaidman2016}, the outcome of the measurement of spins of the second GHZ quantum state along the x-direction defines the choice of either x- or y-direction to measure the spin of the first GHZ quantum state. Both quantum mechanics and the many-worlds interpretation allow for only 16 possible combinations of outcomes for six spin-$\frac{1}{2}$ measurements in four splits of the Multiverse, while a free choice of measurement directions for the first GHZ quantum state leads to 64 possible combinations of the outcomes.

This interpretation of the Bell theorem allows keeping a local hidden variable description of reality without the need of instant and non-local ``spooky action at a distance". However, the theory still has to have non-locality in terms of preserving perfect correlations for spatially separated objects. We argue that the CDFT is an excellent candidate to describe physical reality in a local and deterministic way leading to these perfect spatial correlations within a single split of the Multiverse.

\textbf{Quantum entanglement} still does not have a good explanation beyond the Copenhagen interpretation, which allows for quantum states of spatially separated particles that can be described by a joint wavefunction, but any subset of these particles cannot be described separately from other particles. The proposed theory provides an interpretation of this phenomenon in a local and deterministic way.

We argue that quantum entanglement arises from non-computable propagation rules. Without them, the Multiverse would have all possible configurations of integer value fields and directions of links subject to a unique Big Bang event and the directed acyclic graph property. Non-computable propagation rules set additional global constraints that limit the number of possible local splits of the Multiverse, and therefore create strong long-distance correlations between discrete fields within a single split of the Multiverse.

\textbf{Unitary evolution for quantum fields.} On a final note, we introduce a direction for future research toward a quantum field theory interpretation of the CDFT that will have unitary evolution for quantum fields. We use Gerard `t Hooft's ideas for a deterministic interpretation of quantum mechanics, which follows from the cyclical evolution of discrete systems~\cite{tHooft2016}.

We get back to a stationary Universe example with a nanotube graph to demonstrate unitary evolution. There are microstates $|\Phi_{i}\rangle$ and macrostates $|\Psi_{q}\rangle$ for each charge value $q$. All microstates form a Markov stochastic matrix $\mathcal{M}$ for probabilistic transitions from one microstate to another. This matrix has non-overlapping probability paths for each macrostate with charge $q$.

The key idea is to find a unitary matrix $U$ in the space of abstract vectors $|\Phi_{i}\rangle$, that has the same transition probabilities $P_{jk}=\mathcal{M}_{kj}=|U_{kj}|^2$.

There are two general types of stochastic matrices that have this property. One example is a Hadamard matrix in $N$ dimensions: $U_{jk}=H_{jk}=\exp{[2\pi i(j-1)(k-1)/N]}/\sqrt{N}$. This Hadamard matrix generates probabilistic transitions from each state equally likely to all $N$ states. Another example is a deterministic transition matrix that has all zeros and a single one for each row and each column. The proposed theory with the postulated principle of least action for equiprobable sequential propagation has elements of both Hadamard type evolution and deterministic evolution.

Here is an example of how a particular case of a deterministically evolving CDFT model can have unitary evolution between microstates $|\Phi_{i}\rangle$. We consider deterministic sequential growth, so that there is an algorithm that takes topology of a microstate $|\Phi_{i}\rangle$ and deterministically calculates which next vertex to propagate. If the algorithm is complex enough, then it can generate a pseudo-random sequence, so that the postulated principle of least action still holds statistically. This deterministic propagation uniquely defines a split of the Multiverse, so that any microstate has deterministic future.

Now coming back to the stationary Universe example, all microstates during sequential growth will have a unique immediate descendant. Some microstates will form circles of consecutive transitions, and some microstates will converge to these circles exactly as Gerard `t Hooft proposed in his work on the deterministic interpretation of quantum mechanics~\cite{tHooft2016}. From the analogy in real physics, microstates from all circles form true vacua, and the remaining microstates form false vacua that eventually decay to one of the true vacua.

Following the ideas in~\cite{tHooft2016}, circular evolution of microstates in a certain true vacuum is described by an evolution operator with a unitary matrix $U_{op}$ that has some Hermitean operator with a Hermitean matrix $H_{op}$:

\begin{equation}\label{eq:unitary_evolution}
U_{op} = e^{-i H_{op} \delta t}
\end{equation}

That is how a Hilbert space for the quantum mechanical description may appear in the proposed theory. Microstates $|\Phi_{i}\rangle$ are ontological states in this Hilbert space that allows for a change of the basis to any linear superposition of them. This approach shows the way how complex numbers appear in the proposed theory, which is essential because they are crucial for quantum mechanical description.

A static Universe with deterministic propagation is a minimal case. More research is required to study the relevance of general CDFT models for the interpretation of quantum mechanics. Two key questions are what to do with an expanding Universe and whether deterministic propagation is a crucial part of the theory.

\section{Conclusion}

We propose a novel approach to quantum gravity to unite general relativity and quantum mechanics. We start with a core postulate that a complete description of physical reality has to be causal, discrete, local, and deterministic. We propose to consider models with fundamental symmetries of isotropy, CPT invariance, and charge conservation. The resulting theory is a covariant combination of cellular automata and the causal set theory on self-similar graphs. The beauty of this theory is that it is static. In some sense, the Universe has already happened, and time is just an illusion that follows from causal relations for the fields around an observer.

We studied only simple computable models for hundreds of Planck epochs from the Big Bang because the computation of even a hundred epochs in four dimensions takes weeks on a single computer. The theory reproduces an eternally expanding closed Universe with a physically relevant scale for the Hubble constant and the cosmological term. We only speculated on how non-computable models can be relevant to demonstrate mass and inertia, and to reproduce cosmological parameters for an early Universe then matter played a dominant role in spacetime expansion. We also demonstrated unitary evolution only for a stationary Universe with deterministic propagation. These are two major week areas for the proposed theory that require additional research in the future.

Any scientific theory has to be falsifiable and to have an experimental confirmation. There are several approaches to that. First, one may try to explain large-scale phenomena, e.g., the patterns of discrete fields near the Big Bang event may explain the evolution of cosmological variables for early stages of expansion of the Universe~\cite{Overduin1998}, or the underlying structure of the self-similar graph may be relevant to explain anisotropy and angular differences in higher moments of the cosmic microwave background radiation~\cite{White1994}.  Second, one may perform an ab initio calculation of known fundamental constants, e.g., by defining energy from quantum field theory approach and checking its correspondence with gravitational energy derived from geometry. Finally, one may search for stable patterns that may represent light or elementary particles and study interactions between them to get to the Standard Model of particle physics as low energy and long-distance approximation. 

There is still much room to prove rigorously that the proposed causal discrete field theory is relevant to describe physical reality. If this is a right approach for quantum gravity, there is still a need to find the right combination of two things: an underlying self-similar graph and a propagation rule. In any case, the proposed theory opens up a novel way to describe the Universe at all scales.

\bibliography{CDFT}

\begin{thebibliography}{55}
\bibitem [{mat()}]{matlab}%
  \BibitemOpen
  \href@noop {} {}\bibinfo {note} {The Matlab code is available by
  request}\BibitemShut {NoStop}%
\end{thebibliography}

\end{document}